\documentclass{svjour3}                     
\smartqed  
\usepackage{graphicx}
\begin{document}

\title{Tidal interactions in multi-planet systems
}

\titlerunning{Tides and multi-planet systems}        

\author{J.C.B. Papaloizou        
}

\authorrunning{Papaloizou} 

\institute{J.C.B.  Papaloizou \at
              DAMTP, Centre for Mathematical Sciences,\\
               Wilberforce Road, Cambridge CB3 0WA, UK\\
              Tel.: +44 1223 765000\\
              Fax: +44 1223 765900\\
              \email{jcbp2@damtp.cam.ac.uk}           
}

\date{Received: date / Accepted: date}

\maketitle

\begin{abstract}
We  study 
  systems of close orbiting planets evolving under the influence
of tidal circularization.
It is supposed that a  commensurability forms   through the action of 
disk induced migration and orbital circularization. After the system enters an inner cavity
or the disk disperses the evolution continues under the influence of tides due to
the central star which induce orbital  circularization.
 We derive approximate analytic models  that describe  the evolution away
from a general first order resonance that results from tidal circularization  in  a two planet system
 and which can be shown to be a direct consequence
of the conservation of energy and angular momentum. We consider 
the situation when the system is initially very close to resonance 
and  also  when the system is between resonances.
 We also perform numerical simulations which confirm these models and then apply  them to
two and four planet systems chosen to have  parameters related to  the GJ581 and HD10180 systems.
We also estimate the tidal dissipation rates through effective quality factors 
 that could result in   evolution to observed period ratios within the 
lifetimes of the systems.  
 Thus the survival of, or degree of departure from,  close  commensurabilities  in   observed systems
may be indicative of the effectiveness of tidal disipation,  a feature which in turn may be related to
the internal structure of the planets involved.
\end{abstract}

\section{Introduction}
\label{intro}
Planetary systems  containing  hot Neptunes  and  hot super--Earths  have been observed recently.   
A  system   of this kind consists of the  four planets
around the M--dwarf GJ~581 (Bonfils et al. 2005, Udry et al. 2007, Mayor et al. 2009a).  The projected masses of the planets are 1.9, 15.6, 5.4 and 7.1~M$_{\oplus}$ and the periods are  3.15, 5.37, 12.93 and 66.8 days, respectively.
 Other such multiple systems   are  that around  HD~40307 (Mayor et al. 2009b) which consists of  three  planets with projected masses of 4.2, 6.9 and 9.1~M$_{\oplus}$ and periods of  4.31, 9.62 and 20.46 days, respectively
and that around HD 10180 (Lovis et al. 2010) which consists of seven planets with projected masses 
1.35, 13.10,  11.75,  25.1,  23.9,  21.4  and    64.4~M$_{\oplus}$
and periods of  1.18,   5.76, 16.36, 49.74, 122.76,  601.2, and   2222 days respectively. The innermost member of the latter system
has yet to be confirmed.

Migration due to tidal interaction with the disk is a  possible  mechanism through  which
planets end up on short  period orbits, as {\em in situ} formation implies  very massive discs (e.g., Raymond et al. 2008).  
Terquem \& Papaloizou 2007 (see also  Brunini \& Cionco 2005)  proposed a scenario for
forming hot super--Earths in which
a population of cores that formed  at some distance from the central star migrated inwards due
to interaction with the disk.  These collided and  merged as they  went~. This process  could produce
 systems of planets with masses in the earth mass range,  located inside an assumed  disk inner edge,
   on short period orbits with mean motions  of neighbouring planets that frequently
exhibited near commensurabilities.
However,  tidal circularization of the orbits induced by tidal interaction with the central star,
together with later close scatterings and  mergers tended to
cause the system to move away from earlier established commensurabilities  to an extent
determined by the effectiveness of these processes. 

Papaloizou \&  Terquem (2010) considered  the system around HD~40307
for which the pairs consisting of the innermost and middle  planets
and the middle and outermost planets are near but not very close to a  pair of 2:1 resonances.
In spite of this it   was found that   secular effects produced by the action of the resonant angles
coupled with the action of tides from the central star could cause the system to increasingly separate from commensurability.
Resonant effects can arise even when departures from strict commensurability
are apparently large because  tidal circularization produces small eccentricities which, for first order resonances,  can be consistent with resonant angle libration  (see Murray \& Dermott 1999).

In this paper we undertake a  further study of
  systems of close orbiting planets evolving under the influence
of tidal circularization. We present simple analytic models  describing the evolution away
from a general first order resonance for a two planet
 system under the influence of tidal circularization,
describing the situation 
both when the commensurabilty is very close
and  also  when the system is between resonances. We also perform numerical simulations
of two and four planet systems chosen to have  parameters related to  the GJ581 and HD10180 systems~.
We consider the situation when various commensurabilities result through the action of assumed
disk induced migration and orbital circularization rates,
estimating  the magnitudes of tidal quality factors that could produce  evolution to observed period ratios within the 
lifetimes of the systems.

The plan of this paper is as follows.
In  section\ref{simple1} we consider  a coplanar system of  planets in near circular orbits
in which orbital  energy is dissipated while its total angular momentum is conserved.
The  system is expected to spread  in a  similar manner  to  a viscous accretion disk 
(see Lynden-Bell \& Pringle 1974).  For a two planet system, this radial  spreading
will always  lead to evolution away from  an  initially  close  commensurability.
 
 In sections \ref{coords}, \ref{disk tides}  and  \ref{p+1p}, we 
carry out an anlaytic study
of two planets near to a first order commensurability  under the influence of tidal circularization.
 For planets in the mass range we consider, it is readily estimated
that tides raised on the planet are very much more important than tides 
raised on the star (eg. Goldreich \& Soter 1966, Barnes et al.  2009).
 In addition  the orbital decay timescale due to
tides raised on  the star
may be   estimated to be  much longer than
any timescale of interest (eg. Barnes et al. 2009).
Thus  tides raised on the star have  been  neglected.

We consider the initial evolution away from a  close first order  
commensurability in section  \ref{tight}
and go on to consider the case of evolution when the commensurability 
 is not so close, or the system is 
 between commensurabilities in section \ref{loose}~.
 As expected from the simple arguments  given in section \ref{simple1},
the system departs from an initially close 
commensurability  moving to a neighbouring one.

We go on to
perform   numerical simulations of multiplanet systems  in section \ref{Numsim}.
Various commensurabilities between pairs of planets are set up by applying dissipative forces
assumed to arise from a disk, that lead to orbital migration and circularization.
These forces were then removed corresponding to assumptions of either entry into an inner cavity
or removal of the disk. The evolution of the system under tidal circularization caused by interaction
with the central star was then followed. For illustrative purposes we consider  two planet 
systems  with  parameters corresponding to the two innermost planets in the  GJ581 system.
In section \ref{3:2} we consider a system  that formed  a 3:2 commensurability
which then evolved under orbital circularization indicating that the
model system  could attain the period ratio appropriate to  the actual system  if the tidal
parameter $Q'$  introduced by Goldreich \& Soter (1966)   $ \sim 100.$
The situation when the system began with disk parameters that led to 
 a 5:3 commensurability is then  similarly  studied in section \ref{5:3}.
We go on to consider the effect of  adding the additional planets in the GJ581 
system  in section \ref{extraplanets}.

 As there are examples of low mass planetary systems
such as  HD 10180 which have separations of pairs of planets, the third and fourth innermost in that case,
 that indicate there may have been  a past proximity to a 3:1 commensurability, 
  we consider  an  exploratory simulation  of a system 
for which the initial disk evolution  sets up  a 3:1 commensurability  in section \ref{3:1}.
Finally in section \ref{Discuss} we summarize and discuss our results.

%



%
%

\section{Commensurabilities and  tidal circularization in planetary systems }\label{simple}
We begin by considering the evolution of planetary systems undergoing tidal circularization
in a general way and then  move on to consider simple analytic models of two planet systems 
 that can be close to first order commensurabilities.

\subsection{Two interacting planets in circular orbits for which energy is dissipated at fixed total angular momentum}\label{simple1}
Consider two interacting planets with orbital energies $E_1$ and $E_2$ respectively.
The associated orbital angular momenta for assumed circular orbits are $-2E_1/n_1$ and $-2E_2/n_2$
respectively. Here $n_1$ and $n_2$ are the mean motions
associated with the two planets. Suppose now the system dissipates energy while conserving its total angular momentum.
This is expected to be the case during orbital circularization when this occurs as a result of 
stellar tides dissipated in the planets because the planets themselves cannot contain a significant 
amount of angular momentum.
Accordingly we write
\begin{eqnarray}
\frac{d E_1}{dt}+  \frac{d E_2}{dt}&=&-{\cal{L}},
\end{eqnarray}
where ${\cal{L}}$ is the rate of energy dissipation.
Angular momentum conservation implies that
\begin{eqnarray}
  \frac{1}{n_1}\frac{ d E_1}{dt} &= &- \frac{1}{n_2}\frac{ d E_2}{dt} 
 \end{eqnarray}
from which we obtain 
\begin{eqnarray}
\frac{d E_1}{dt}&=&-\frac{{\cal{L}}}{1-n_2/n_1  },\label{simple3}
\end{eqnarray}
and
\begin{eqnarray}
\frac{d E_2}{dt}&=&-\frac{{\cal{L}}}{1-n_1/n_2  }.
\end{eqnarray}
Supposing  that $n_1 > n_2,$ the above two equations 
imply that planet $1$ moves inwards losing energy
while planet $2$ moves outwards,  taking up the angular momentum lost by 
planet $1.$
This is the generic form for the evolution of an accretion disc (see Lynden-Bell  \& Pringle 1974).

We now go on to discuss some simplified models for the interaction  of two planets
that may be either very close to or some distance away  from a strict  first order commensurability.
In these contexts we show how tidal dissipation induced by forced eccentrcities
causes the system to separate. When the resonance  is close,  this causes the system to depart
further from commensurability. 

\subsection{Coordinate system}\label{coords}
We consider  a general system of $N$ planets orbiting a central mass.
We adopt Jacobi coordinates (Sinclair~1975, Papaloizou \& Szuszkiewicz~2005) for which the radius vector
 of  planet $i,$  ${\bf r }_i,$ is measured relative
to the  centre of mass of  the system comprised of a dominant central
mass  $M$ and  all other  planets
interior to  $i,$ for  $i=1,2...,N.$    The planets are assumed to
maintain an ordering  with increasing   $i$ corresponding to greater
distances $|{\bf r }_i|$ from the dominant central mass.   Thus the
innermost planet has $i=1.$ 
The  Hamiltonian,  correct to second order in the planetary masses,  can be written in the form:
\begin{eqnarray} H & = &  \sum_{i=1}^N \left({1\over 2}  m_i | \dot {\bf r}_i|^2
- {GM_{i}m_i\over  | {\bf r}_i|} \right)   \nonumber \\
& - &\sum_{i=1}^{N-1}\sum_{j=i+1}^NGm_{i}m_j
\left({1 \over  | {\bf r}_{ij}|}  -  { {\bf r}_i\cdot {\bf r}_j
\over  | {\bf r}_{j}|^3}\right).
\end{eqnarray}
Here $M_{i}=M+m_i $ and
$ {\bf r}_{ij}= {\bf r}_{i}- {\bf r}_{j}.$

The equations of motion for motion  for  planet $i$ assumed to move  
in a fixed plane, about a dominant central mass,
 may be written in the form
(see, e.g., Papaloizou~2003, Papaloizou \& Szuszkiewicz~2005):

\begin{eqnarray}
\dot E_i &=& -n_i\frac{\partial H}{\partial \lambda_i}\label{eqnmo1}\\
\dot L_i &=& -\left(\frac{\partial H}{\partial \lambda_i}+\frac{\partial H}{\partial \varpi_i}
\right)\label{eqnmo2}\\
\dot \lambda_i &=& \frac{\partial H}{\partial L_i} + n_i \frac{\partial H}{\partial E_i}\\
\dot \varpi_i &=& \frac{\partial H}{\partial L_i}.\label{eqnmo4}
\end{eqnarray}
Here the orbital  angular momentum of  planet  $i$ which has
reduced mass
$m_i = m_{i0}M/(M+m_{i0}),$ with $m_{i0}$ being the actual mass,  is $L_i$ and the
orbital  energy is $E_i.$
For motion around a central point  mass $M$ we have:
\begin{eqnarray}
     L_i &=&  m_{i}\sqrt{GM_{i}a_i(1-e_i^2)}, \\
     E_i &=& -{{GM_{i}m_{i}}\over{2a_i}},
\end{eqnarray}
where $M_{i} = M+m_{i0},$  $a_i$ denotes the semi-major axis and $e_i$  the eccentricity
of planet $i.$

\noindent The  mean longitude of planet $i$ is $\lambda_i = n_i (t-t_{0i}) +\varpi_i ,$
 where $n_i  = \sqrt{GM_{i}/a_i^3}$ is its mean motion, with
$t_{0i}$ denoting its time of periastron passage 
and $\varpi_i$ the longitude of periastron.

\noindent From equations (\ref{eqnmo1}) and (\ref{eqnmo2})
an equation for the evolution of the eccentricity of planet $i$ may be  readily
obtained in the form  
\begin{eqnarray}
\dot e_i &=&
\frac{ \sqrt{1-e_i^2}}{e_i m_i n_i a_i^2}\left(\frac{\partial H}{\partial \lambda_i}\left(\sqrt{1-e_i^2}-1\right)+\frac{\partial H}{\partial \varpi_i}\right)
\label{eqnmo5}.
\end{eqnarray}

\noindent The Hamiltonian may 
 be expanded in a Fourier series
involving linear combinations of the  $(2N-1)$ angular  differences
$\varpi_i -\varpi_1,  i=2,3..N$  and
$\lambda_i - \varpi_i, i=1,2,..N .$ 
In the limit of small eccentricities of interest here,
only terms that are of first  order in the eccentricities need to be retained
(terms that are of zero order do not lead to changes to eccentricities or to resonances).
If this is done  the possibility of  first order resonances, 
for which the ratio of the periods of two planets  is the ratio of successive integers, is allowed for. 
The above  approximation scheme should be  valid when  circularization times are small enough
to ensure that the eccentricities remain small. This situation is realized
for  examples of  low mass protoplanets migrating in protoplanetary discs
(Papaloizou \& Szuszkiewicz 2005).

Near a first order  $p+1 : p $ resonance, $p$ being an integer,  we expect that
terms in the Hamiltonian involving   angles of the type
$\phi_{p,j,i,j} = (p+1)\lambda_j-p\lambda_i-\varpi_j, $ and
$\phi_{p,j,i,i}= (p+1)\lambda_j-p\lambda_i-\varpi_i,$
where the subscripts on the left hand side correspond to those on the
right hand reading from left to right,
will be slowly varying and thus be dominant.  Accordingly we shall
retain only terms of this type. Motion away from resonances may also
be considered having made this approximation although neglected high frequency
modulations may be more significant then.   

The  Hamiltonian   may  be written
in the form
 \begin{equation}H=\sum _{i=1}^N E_i +\sum_{i=1}^{N-1}\sum_{j=i+1}^N H _{i,j},\label{Hamiltonian}
 \end{equation}
 where the component of the interaction Hamiltonian 
  $H_{i,j}$ that is first order in the eccentricities is given, given that $j>i,$  by
\begin{equation} H _{i,j}= -\frac{Gm_im_j}{a_j}\sum_{p=1}^{\infty}\left(  e_j C_{p,j,i,j}\cos (\phi_{p,j,i,j})+e_iD_{p,j,i,i}\cos (\phi_{p,j,i,i}) \right), \label{Hamil} \end{equation}
with
\begin{equation} C _{p,j,i,j}={1 \over 2}\left(   x{d(b^{(p)}_{1/2}(x))\over dx} +(2p+1)b^{(p)}_{1/2}(x)
-4x\delta^{p}_1 \right) \ \ {\rm and} \ \ \label{Hamil1} \end{equation}
\begin{equation} D_{p,j,i,i}= -{1 \over 2}\left(   x{d(b^{(p+1)}_{1/2}(x))\over dx} +2(p+1)b^{(p+1)}_{1/2}
(x)  \right) .\label{Hamil2} \end{equation}
Here $b^{(p)}_{1/2}(x)$ denotes the usual Laplace coefficient
(e.g. Brouwer \& Clemence 1961)
with the argument $x = a_i/a_j$ and $\delta^{p}_1$ denotes
the Kronnecker delta. We remark that the subscripts associated with the coefficients
$C _{p,j,i,j} $ and  $D_{p,j,i,i}$ correspond to the related angles as in (\ref{Hamil}). 
We shall also  make the approximation of replacing
 $M_{i}$ by $M$ and equivalently $m_{i0}$ by $m_{i}.$

The governing equations for motion, retaining only  terms that are of the lowest order in the eccentricities, follow  from Hamilton's equations 
(\ref{eqnmo1})- (\ref{eqnmo4}) for the Hamiltonian (\ref{Hamiltonian})
discussed above  as
\begin{eqnarray}
\frac{d e_i}{dt} &= &
-\sum_{p=1}^{\infty}\left[\sum_{j=i+1}^N\frac{Gm_jD_{p,j,i,i}\sin (\phi_{p,j,i,i})}{n_ia_i^2a_j} 
+\sum_{j=1}^{i-1}\frac{Gm_jC_{p,i,j,i}\sin (\phi_{p,i,j,i})}{n_ia_i^3}\right]\label{eqnei}\\
\frac{d n_i}{dt} &= &
-\sum_{p=1}^{\infty}\left[
\sum_{j=i+1}^N\frac{3Gm_jp}{a_i^2 }\left(\frac{e_iD_{p,j,i,i}\sin (\phi_{p,j,i,i})}{a_j} 
+\frac{e_jC_{p,j,i,j}\sin (\phi_{p,j,i,j})}{a_j}\right) 
\right.\nonumber\\
&-&\left. \sum_{j=1}^{i-1}\frac{3Gm_j(p+1)}{a_i^2 }\left (\frac{e_jD_{p,i,j,j}\sin (\phi_{p,i,j,j})}{a_i}
+\frac{e_iC_{p,i,j,i}\sin (\phi_{p,i,j,i})}{a_i}\right )\right ]
\label{eqnni}\\
\frac{d\varpi_i}{dt}&= &
\sum_{p=1}^{\infty}\left[\sum_{j=i+1}^N\frac{Gm_jD_{p,j,i,i}\cos (\phi_{p,j,i,i})}{e_in_ia_i^2a_j} 
+\sum_{j=1}^{i-1}\frac{Gm_jC_{p,i,j,i}\cos(\phi_{p,i,j,i})}{e_in_ia_i^3 }\right]
.\label{eqnvarpi} \\
\nonumber
\end{eqnarray}
In addition, consistent with the above approximation scheme, the rate of change of the mean longitudes 
may be obtained from

\begin{equation}
\hspace{-9cm}\frac{d \lambda_i}{dt} = n_i
\label{longit}
\end{equation}
which also enables  evaluation of the rate of change of the angles
$\phi_{p,j,i,i},  \phi_{p,j,i,j} $ etc.

\subsection{The incorporation of disk tides }\label{disk tides}
We incorporate     the effects of orbital  circularization 
by adding additional terms to the right
hand sides of  equations (\ref{eqnei}) and (\ref{eqnni}).
Equation (\ref{eqnei})   is  modified through the straightforward prescription
\begin{equation}
\frac{d e_i}{dt} \rightarrow \frac{d e_i}{dt} - \frac{e_i}{t_{c,i}},\label{circi}
\end{equation}
where $t_{c,i}$ is the circularization time for planet $i.$
Similarly equation (\ref{eqnni})   is  modified according to 
\begin{equation}
\frac{d n_i}{dt} \rightarrow \frac{d n_i}{dt} +\frac{3n_ie_i^2}{t_{c,i}}.\label{circi1}
\end{equation}
This adjustment is necessary to
account  for the orbital energy dissipation occurring as a result of circularization
correct to  the lowest order in $e_i.$
This dissipation is assumed to occur with out changing the angular momentum
of the system because the planets can only potentially contain
a negligible amount of angular momentum compared to that in the orbit.
It follows from the energy dissipation rate for planet, $i,$  given by
\begin{equation}
\frac{dE_i}{dt} = - \frac{ m_i n_i^2a_i^2e_i^2}{(1-e^2_i)t_{c,i}}.\label{Edissip}
\end{equation}
For small eccentricities,  $e_i^2$ may be neglected in the denominator  
of the above expression and the total rate of energy dissipation in the system
is obtained by summing over all planets.
   
\section{Two planets in a  $p+1:p$ commensurability}\label{p+1p}
It is possible to investigate solutions of equations (\ref{eqnei}) - (\ref{eqnvarpi})
modified to incorporate circularization, that illustrate the  geometrical separation 
of the system as energy is dissipated while the total angular momentum
is conserved, in a number circumstances. 
\subsection{A tight commensurability}\label{tight}
We begin with an example where
two successive 
 planets $k$ and $k+1$ maintain a  $p+1:p$ commensurability with the associated angles
in a state of at most  small amplitude libration  while their semi-major axes
separate. We later go on to consider a simple restricted example where the angle circulates.
The effects of planets other than the resonant pair is neglected.
Equations(\ref{eqnei}) - (\ref{eqnvarpi}) with the modifications given by
 (\ref{circi}) and (\ref{circi1}) to incorporate circularization give the governing
 equations for planet $k$ in the form
 \newpage
\begin{eqnarray}
\frac{d e_k}{dt} &= &
-\frac{Gm_{k+1}D_{p,k+1,k,k}\sin (\phi_{p,k+1,k,k})}{n_ka_k^2a_{k+1}}
- \frac{e_k}{t_{c,k}} \label{eqnei2}\\
\frac{d n_k}{dt} &= &
-\frac{3Gm_{k+1}p}{a_k^2 }\left(\frac{e_kD_{p,k+1,k,k}\sin (\phi_{p,k+1,k,k})}{a_{k+1}}\right.\nonumber\\ 
  &+&\left.\frac{e_{k+1}C_{p,k+1,k,k+1}\sin (\phi_{p,k+1,k,k+1})}{a_{k+1}}\right) 
+\frac{3n_ke_k^2}{t_{c,k}}\label{eqnni2}\\
\frac{d\varpi_k}{dt}&= &
\frac{Gm_{k+1}D_{p,k+1,k,k}\cos (\phi_{p,k+1,k,k})}{e_kn_ka_k^2 a_{k+1}} 
.\label{eqnvarpi2} 
\end{eqnarray}
Similarly the governing equations for planet $k+1$ are given by
\begin{eqnarray}
\frac{d e_{k+1}}{dt} &= & 
-\frac{Gm_kC_{p,k+1,k,k+1}\sin (\phi_{p,k+1,k,k+1})}{n_{k+1}a_{k+1}^3}
- \frac{e_{k+1}}{t_{c,k+1}}\label{eqne21}\\
\frac{d n_{k+1}}{dt} &= &
\frac{3Gm_k(p+1)}{a_{k+1}^2 }\left (\frac{e_kD_{p,k+1,k,k}\sin (\phi_{p,k+1,k,k})}{a_{k+1}}\right.\nonumber\\
 &+&\left.\frac{e_{k+1}C_{p,k+1,k,k+1}\sin (\phi_{p,k+1,k,k+1})}{a_{k+1}}\right )
+\frac{3n_{k+1}e_{k+1}^2}{t_{c,{k+1}}}\label{eqnni21}\\
\frac{d\varpi_{k+1}}{dt}&= &
\frac{Gm_kC_{p,k+1,k,k+1}\cos(\phi_{p,k+1,k,k+1})}{e_{k+1}n_{k+1}a_{k+1}^3 }.\label{eqnvarpi21} 
\end{eqnarray}
Setting $\phi_{p,k+1,k,k}\rightarrow \phi_{p,k+1,k,k}\pm\Delta \phi_{p,k+1,k,k}$
and  $\phi_{p,k+1,k,k+1}\rightarrow \phi_{p,k+1,k,k+1}\pm\Delta \phi_{p,k+1,k,k+1},$
where the  positive sign is taken when 
the  equilibrium value of the  angle is zero and the negative sign
is taken when it is $\pi,$ and  $\Delta$ indicates a  small shift such that the sines  of the angles
may be replaced by the angles themselves.
Then assuming that the evolutionary time scale  is much longer than the
circularization times so that the time derivatives of the eccentricities
may be neglected,   we can then find expressions for   the small angular  shifts in the form 
\begin{eqnarray}
\Delta \phi_{p,k+1,k,k}&=&-\frac{e_kn_ka_k^2a_{k+1}}{Gm_{k+1}D_{p,k+1,k,k}t_{c,k}}.\\
\Delta \phi_{p,k+1,k,k+1}&=&-\frac{e_{k+1}n_{k+1}a_{k+1}^3}{Gm_{k}C_{p,k+1,k,k+1}t_{c,k+1}}.
\end{eqnarray}
Substituting these into  the equations for the evolution of the mean motions  yields
\begin{eqnarray}
\frac{d n_{k}}{dt} &=&\frac{3(p+1)e_{k}^2n_{k}}{t_{c,k}}+
\frac{3pm_{k+1}e_{k+1}^2n_{k+1}a_{k+1}^2}{m_ka_k^2t_{c,k+1}}\\
\frac{d n_{k+1}}{dt} &=&-\frac{3(p+1)m_ke_{k}^2n_{k}a_k^2}{m_{k+1}t_{c,k}a_{k+1}^2}-
\frac{3pe_{k+1}^2n_{k+1}}
{t_{c,k+1}}.
\end{eqnarray}
The above pair of equations express the conservation of energy and angular 
momentum for the system in the limit of small eccentricity.  We remark that 
the latter follows in  the form
\begin{equation}
m_{k+1}a_{k+1}^2\frac{d n_{k+1}}{dt} +m_ka_k^2\frac{d n_{k}}{dt}=0,
\end{equation}
while the former follows from using the fact that $E_j \propto m_j n_j^{2/3}$
to find equations for $dE_j/dt,  j=k,k+1$ and then adding.
We may also obtain an equation showing how the period ratio
increases with time in the form 
\begin{equation}
\frac{d}{dt}\left(\frac{n_k}{n_{k+1}}\right)=\frac{3n_k J}{n_{k+1}}
\left[\frac{(p+1)e_k^2}{t_{c,k}J_{k+1}}
+\frac{pe_{k+1}^2}{t_{c,k+1}J_{k}}\right],\label{Psep}
\end{equation}
where $J_k =m_ka_k^2n_k,$ and $J=J_{k}+J_{k+1}.$
In order to proceed further we need to calculate the eccentricities.
These may be obtained from the governing equations for the evolution 
of the angles  that may be obtained from (\ref{longit}), (\ref{eqnvarpi2}) and (\ref{eqnvarpi21}) in the form
\begin{eqnarray}
\frac{d \phi_{p,k+1,k,k}}{dt}&=&(p+1)n_{k+1}-pn_k- \frac{Gm_{k+1}D_{p,k+1,k,k}
\cos (\phi_{p,k+1,k,k})}{e_kn_ka_k^2a_{k+1}} .\\
\frac{d\phi_{p,k+1,k,k+1}}{dt}&=&(p+1)n_{k+1}-pn_k- 
\frac{Gm_kC_{p,k+1,k,k+1}\cos(\phi_{p,k+1,k,k+1})}{e_{k+1}n_{k+1}a_{k+1}^3}.
\end{eqnarray}

As the angles are quasi-steady and close to  zero or $\pi,$ these  expressions enable the calculation of the  squares of the 
eccentricities $e_k$ and $e_{k+1}$ which are  required in order
 to calculate the rate of period separation through (\ref{Psep}). They are found to be given by
\begin{eqnarray}
e_k^2&=&\left( \frac{Gm_{k+1}D_{p,k+1,k,k}}{n_ka_k^2a_{k+1}
 [(p+1)n_{k+1}-pn_k]}\right)^2\hspace{1cm}  {\rm and}\label{ecclose1}\\
e_{k+1}^2&=&
\left(\frac{Gm_kC_{p,k+1,k,k+1}}{n_{k+1}a_{k+1}^3[(p+1)n_{k+1}-pn_k]}\right)^2
\hspace{1cm} {\rm respectively}\label{ecclose2}.
\end{eqnarray}
Using these in (\ref{Psep}) we obtain
\begin{equation}
\frac{d}{dt}\left(\frac{n_k}{n_{k+1}}-\frac{p+1}{p}\right)^3=\frac{9n_k J}{n_{k+1}}F,\label{sepeq}
\end{equation}
where
\begin{equation}
F=\frac{(p+1)}{t_{c,k}J_{k+1}}\left(\frac{Gm_{k+1}D_{p,k+1,k,k}}{pn_kn_{k+1}a_k^2a_{k+1}}\right)^2
+\frac{p}{t_{c,k+1}J_{k}}\left(\frac{Gm_kC_{p,k+1,k,k+1}}{pn_{k+1}^2a_{k+1}^3}\right)^2.
\end{equation}
When the system starts to move away from a  commensurability taken to be exact
at $t=0,$  we may treat the right hand side of
(\ref{sepeq}) as being constant and integrate with respect to time to obtain
\begin{equation}
\frac{n_k}{n_{k+1}}-\frac{p+1}{p}=\left(\frac{9n_k J}{n_{k+1}}Ft\right)^{1/3}.\label{sepeqt}
\end{equation}
A similar scaling for which the separation from a commensurability increases $\propto t^{1/3}$
was obtained for a three planet system by Papaloizou \& Terquem (2010).

\subsection{The interaction between two planets away from a close commensurability }\label{loose}
In this case we again assume  interaction between planets
$k$ and $k+1.$ In this case we consider the situation away from a strict commensurability
where significant libration or circulation may occur.
This is a natural development as tidal evolution  causes the system
to evolve away from a tight commensurability of the type described above towards
such a situation.
We make the additional simplification of assuming that 
$m_{k+1} \gg m_{k}.$ In that case, a  circular restricted $3$ body
problem  may be adopted.  Only the  motion of planet $k$
is considered with $e_{k+1}=0.$  Equations (\ref{eqnei}) - (\ref{longit})
apply and as $e_{k+1}=0,$  only    terms involving the  angles 
 $ \phi_{p,k+1,k,k}, p=1,2...$
appear. These give the  equations governing the evolution as
\begin{eqnarray}
\frac{d e_k}{dt} &= &- \frac{e_k}{t_{c,k}}
-\sum_{p=1}^{\infty}\frac{Gm_{k+1}D_{p,k+1,k,k}\sin (\phi_{p,k+1,k,k})}{n_ka_k^2a_{k+1}}
\label{eqnei3}\\
\frac{d n_k}{dt} &= &\frac{3n_ke_k^2}{t_{c,k}}
-\sum_{p=1}^{\infty}\frac{3Gm_{k+1}pe_kD_{p,k+1,k,k}\sin (\phi_{p,k+1,k,k})}{a_{k+1}a_k^2 } 
\label{eqnni3}\\
\frac{d \phi_{r,k+1,k,k}}{dt} &=& (r+1)n_{k+1}-rn_k - 
\sum_{p=1}^{\infty}\frac{Gm_{k+1}D_{p,k+1,k,k}\cos (\phi_{p,k+1,k,k})}{e_kn_ka_k^2 a_{k+1}} ,\nonumber\\
 r=1,2, 3& .& \hspace{0mm} .\hspace{2mm} .\label{eqnvarpi3} 
\end{eqnarray}

Although we consider the effect of more than one angle, we focus on a particular
one with $r=q$ which might be considered to be the one closest to
resonance, though that is not essential.  
Setting $x=e_k\cos (\phi_{q,k+1,k,k})$ and $y=e_k\sin (\phi_{q,k+1,k,k})$
in equations (\ref{eqnei3})~and~(~\ref{eqnvarpi3}~)   leads to a  system
that, unlike the original one, 
does not contain an apparent singularity as $e_k\rightarrow 0$ in the form
\begin{eqnarray}
\frac{d x}{dt} &= &
-\omega_q y 
- \frac{x}{t_{c,k}} -
\sum_{p=1\ne q}^{\infty}\alpha_p\sin[(p-q)(\lambda_{k+1}-\lambda_{k})] \label{eqnei4}\\
\frac{d y}{dt} &=& -\alpha_q+ 
\omega_q x
- \frac{y}{t_{c,k}}-\sum_{p=1\ne q}^{\infty}\alpha_p\cos[(p-q)(\lambda_{k+1}-\lambda_{k})] ,
\label{eqnvarpi4} 
\end{eqnarray}
where, recalling that $k$ is fixed, we define
 $\alpha_q=Gm_{k+1}D_{q,k+1,k,k}/(n_{k}a_{k}^2a_{k+1})$\newline
and $\omega_q=(q+1)n_{k+1}-qn_{k}.$

We now remark that if we consider the limit $e_k\rightarrow 0,$ 
which occurs far enough away from resonance, we may neglect the evolution of $n_{k}$
and asume that it  remains constant and equal to $n_{k0}.$
To see this it follows from (\ref{eqnei3}) that $e_k$ scales as $m_{k+1}$
while (\ref{eqnni3}) then indicates that  the change in  $n_k,$
$\delta n_k=n_{k}- n_{k0}$ scales as $m_{k+1}^2$ or $e_k^2 .$
Accordingly, for $q$ of order unity,
 the induced variation of $w_q,$ $\delta \omega_q,$ is such that
$\delta \omega_q \sim n_{k0}e_k^2.$
 By comparing the variation
of the first two terms on the right hand side of
equation (\ref{eqnvarpi4}), it readily follows in   the low eccentricity 
limit  that provided
$ n_{k0}e_k^3 \ll \alpha_q$ or
\begin{eqnarray}
 e_k \ll
\left(\frac{Gm_{k+1}D_{q,k+1,k,k}}{n_{k}^2a_k^2 a_{k+1}}\right)^{1/3},
\end{eqnarray}
the variation of $n_k$ may be neglected so that it may be taken to be
equal to $n_{k0}.$
Similarly $a_k$ is replaced by the corresponding fixed value $a_{k0}.$
The above approximation scheme applies in the low eccentricity limit
or sufficiently far away from strict commensurability such that
$|\omega_q|=|(q+1)n_{k+1}-qn_{k0}|\gg  n_{k0}^{1/3} \alpha_q^{2/3} .$

Given that in the same approximation  (\ref{longit}) implies that

\noindent $\lambda_{k+1}-\lambda_{k}=(n_{k+1}-n_{k0})t,$
equations (\ref{eqnei4}) and (\ref{eqnvarpi4}) describe a linear system
with prescribed harmonic forcing that is easily  solved exactly.
The solution in the limit $t_c \rightarrow \infty$ may be written

\begin{eqnarray}
 x &= &x_0
+\sum_{p=1\ne q}^{\infty}\frac{\alpha_p\cos[(p-q)(\lambda_{k+1}-\lambda_{k})]}
{(p+1)n_{k+1}-pn_{k}} \label{solx}\\
 y &=&-
\sum_{p=1\ne q}^{\infty}\frac{\alpha_p\sin[(p-q)(\lambda_{k+1}-\lambda_{k})]}
{(p+1)n_{k+1}-pn_{k}} ,
\label{soly}
\end{eqnarray}
where $x_0=\alpha_q/\omega_q.$ 
This indicates oscillation about the mean value of $x=x_0.$
In the absence of the periodic forcing circularization would cause
the solution to approach $x=x_0,y=0$ corresponding to a precise
commensurability with zero libration amplitude.
When the forcing is present, there is either libration or circulation depending
on the ratio of the forcing amplitude to $x_0.$ When  $q+1:q$  is the
closest commensurability, this can be small resulting in small amplitude libration.
When some other commensurability is dominant, the motion in the $(x,y)$
plane is  around an approximately    circular curve 
 that encloses the origin and so corresponds to circulation.
 Thus as the system moves through a commensurabilty
 the motion is expected to change from  circulation to libration to circulation
 (see Terquem \& Papaloizou  2007  for an example of  evolution 
 away from a first order commensurability driven by orbital circularization).
Note that in all of these cases the time dependent averages of quantities such as 
$e_k\cos (\phi_{q,k+1,k,k})\equiv x$ and $\cos (\phi_{q,k+1,k,k})$
are generally non zero (see also Papaloizou \& Terquem 2010).

 In order to calculate the rate of energy dissipation resulting from
 orbital circularization and hence the rate of evolution it causes, we
 require the time average of the square of the eccentricity.
 This is given by
 \begin{eqnarray}
\langle e_k^2\rangle=\langle x^2 +y^2 \rangle= \sum_{p=1}^{\infty}\frac{\alpha_p^2}
{[(p+1)n_{k+1}-pn_{k}]^2}.\label{aveesq} 
\end{eqnarray}
We remark that this expression connects to that found
for the tight resonance example (\ref{ecclose1}). The latter expression
is identical to that given by (\ref{aveesq}) if only one term is retained that
corresponds to the tight commensurability considered. Thus we expect the evolution 
to continue to be dominated by the closest comensurability until another
becomes closer and takes over governing the evolution.
\subsection{The orbital evolution of the planet}\label{orbevol}
The rate of change of the orbital energy may be obtained from consideration of (\ref{aveesq})
and (\ref{Edissip}) together with the discussion leading to
equation (\ref{simple3}) given in section \ref{simple1} with the result that
\begin{equation}
\frac{dE_k}{dt} \equiv -\frac{m_ka_k^2n_kn_{k+1}}{3}\frac{d}{dt}\left(\frac{n_k}{n_{k+1}}\right) = - \frac{ m_k n_k^2a_k^2}{(1-n_{k+1}/n_{k})t_{c,k}}
 \sum_{p=1}^{\infty}\frac{\alpha_p^2}
{[(p+1)n_{k+1}-pn_{k}]^2}\label{Edissipk}
\end{equation}
which means that the orbit of $m_k$ contracts and separates from that of $m_{k+1}.$
In the limit $m_{k}/m_{k+1} \rightarrow  0$ in which $n_{k+1}$ becomes fixed,  and close to a commensurability (\ref{Edissipk})
becomes equivalent to (\ref{sepeq}).  Thus  it enables the discussion of the situation
corresponding to a tight commensurability  to be  extended
to conditions away from close commensurability.  
In view if the fact that $n_k/n_{k+1}$ must always increase, such a discussion leads to the conclusion 
that the evolution will be controlled
by successive first order comensurabilities as the system widens (see also Terquem \& Papaloizou 2007).

\section{Numerical Simulations}\label{Numsim}
We here describe simulations of model planetary systems in which commensurabilties
have been formed subsequently evolving under the influence of circularization tides. 
\subsection{Model and initial conditions}\label{siminit}
\label{sec:model}
We consider a
primary star together with
 N~planets embedded in a gaseous disk surrounding it.
The planets undergo gravitational interaction with each other and the
star and are acted on by tidal forces from the disk  and star.
The system is solved as  an  $N$--body  problem.
Tidal interactions are incorporated by applying appropriate
dissipative forces
(see Terquem \& Papaloizou 2007 and Papaloizou \& Terquem 2010 for more details and  examples).
The equations of motion  may be written as
\begin{equation}
{d^2 {\bf r}_i\over dt^2} = -{GM{\bf r}_i\over |{\bf r}_i|^3}
-\sum_{j=1\ne i}^N {Gm_{j0}  \left({\bf r}_i-{\bf r}_j \right) \over |{\bf
    r}_i-{\bf r}_j |^3} -{\bf \Gamma} +{\bf \Gamma}_{i} +{\bf \Gamma}_{r} \; ,
\label{emot}
\end{equation}

\noindent where $M$, $m_{j0}$ and ${\bf r}_j$ denote the mass of
the central star, that of planet~$j$ and the position vector of planet
$j$, respectively.  The acceleration of the coordinate system based on
the central star (indirect term) is given by
\begin{equation}
{\bf \Gamma}= \sum_{j=1}^N {Gm_{j0}{\bf r}_{j} \over |{\bf r}_{j}|^3},
\label{indt}
\end{equation}
\noindent and that due to tidal interaction with the disk and/or the
star is dealt with through the addition of  dissipative forces
(see Papaloizou~\& Larwood 2000). Thus
\begin{equation}
{\bf \Gamma}_{i} = -\frac{1}{t_{mg,i}} \frac{d {\bf r}_i}{dt} -
\frac{2}{|{\bf r}_i|^2 t_{e,i}} \left( \frac{d {\bf r}_i}{dt} \cdot
{\bf r}_i \right) {\bf r}_i - \frac{2}{ t_{i,i}}
\left( \frac{d {\bf r}_i}{dt} \cdot {\bf e}_z \right) {\bf e}_z,
\label{Gammai}
\end{equation}

\noindent where $t_{mg,i}$, $t_{e,i}$ and $t_{i,i}$ are the timescales
over which, respectively, the angular momentum, the eccentricity and
the inclination with respect to the unit normal ${\bf e}_z$ to the  assumed fixed gas
disk midplane change.  Evolution of the angular momentum and
inclination is assumed to be  due to tidal interaction with the disk, whereas
evolution of the eccentricity is assumed to  occur due to both tidal interaction
with the disk and the star.  We have:
\begin{equation}
\frac{1}{t_{e,i}} = \frac{1}{t_{c,i}^d} + \frac{1}{t_{c,i}} ,
\end{equation}
\noindent where $t_{c,i}^d$ and $t_{c,i}$ are the contribution from
the disk and tides raised by the star, respectively.  Relativistic
effects are modeled through ${\bf \Gamma}_{r}$ ( see Papaloizou \&
Terquem 2001).

Because a low mass planet  cannot contain a  significant quantity of
angular momentum,  tides raised on it  by interaction with the star
are assumed  not to  modify the
angular momentum of the  orbit.
We remark that  the orbital decay timescale, due to
tides raised on  the star,
is readily  estimated to be  much longer than
any timescale of interest (eg. Barnes et al 2009)
 thus these tides are ignored from now on.

\subsection {Orbital circularization due to tides from the central star}
\label{sec:startide}

The circularization timescale due to tidal interaction with the star,
in the small eccentricity limit appropriate here, is taken to be
( Goldreich~\& Soter 1966)
\begin{equation}
t_{c,i} = 4.65 \times 10^{4} \; \left( \frac{{\rm M}_\odot}{M}\right)^{3/2}\left( \frac{{\rm M}_\oplus}{m_{i0}}
\right)^{2/3}  \left( \frac{20 a_i}{{\rm 1~au}} \right)^{6.5}  Q'
\; \; \; {\rm years} ,
\label{teccs}
\end{equation}
\noindent where $a_i$ is the semi--major axis of planet~$i.$ Here we
have adopted a mass density of 1~g~cm$^3$ for the planets
(uncertainties in this quantity could be incorporated into a redefinition of $Q'$).
 The parameter
$Q'= 3Q/(2k_2),$ where $Q$ is the tidal dissipation function and $k_2$
is the Love number.
 For solar system planets in the terrestrial mass
range, Goldreich \& Soter (1966) give estimates for $Q$ in the range
10--500 and $k_2 \sim 0.3$, which correspond to $Q'$ in the range
50--2500.  We remark that  this parameter should  be
regarded as being very uncertain for extrasolar planets.
As computations with  increasing $Q'$
become prohibitive on account of
long evolution times,
we have considered values of  $Q'$ of $1.5$ and $3$ in this paper.
However, we have obtained scaling relations which  indicate how to
scale results to larger $Q'.$

\subsection{Type~I migration}\label{TypeImigration}

\begin{figure*}
 \includegraphics[width=\textwidth]{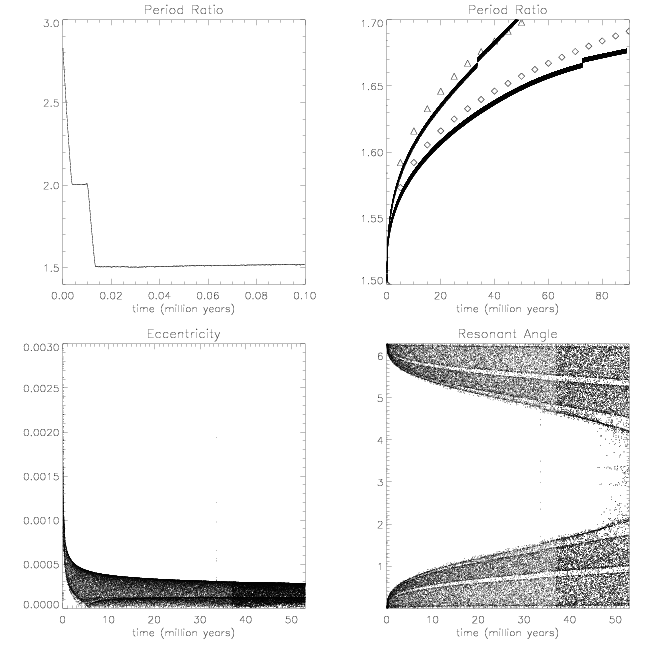}
\caption{ The evolution of two planets that form a 3:2 commensurability.
The early evolution of the period ratio during  convergent migration
is shown in the upper left panel.  The upper right panel shows the
evolution of the period ratio under orbital circularization
after disk migration ceases. The uppermost curve is for $Q'=1.5$ and the
lower curve is for $Q'=3.$ The triangles/diamonds correspond to the analytic predictions
made from equation (\ref{sepeqt}) adapted to the  case of a 3:2 commensurability
for $Q'=1.5/Q'=3$ respectively.
 The evolution of the eccentricity of the
outermost planet is plotted in the lower left panel for $Q'=1.5.$ The evolution
of the resonant angle  $3\lambda_2-2\lambda_1-\varpi_1$
is plotted in the lower right panel for $Q'=1.5.$}
\label{fig:1}       
\end{figure*}

When  a planet is in contact with the
disk, disk--planet interactions occur leading to orbital migration as
well as eccentricity and inclination damping (e.g., Ward~1997).
However,  the  migration rates
to be used are uncertain  even when the disk  surface density
is known,  largely because of uncertainties regarding the
effectiveness of coorbital torques (e.g., Paardekooper \& Melema 2006,
Pardekooper \& Papaloizou 2008, 2009).
 In this context there are indications from modelling the observational data
that the adopted type I migration rate should be significantly
below that predicted by 
the linear calculations of Tanaka et al. (2002)
(see Schlaufman et al 2009).
Hence
we have  carried out simulations with
$t_{mg,i}$ and $t_{e,i}^d$ for any system taken, as for type I migration,  to be proportional to $1/m_i$
 and adopted $t_{i,i}=t_{e,i}^d.$
 A  range of scaling constants  was explored.  These are quoted
together with corresponding numerical results below.

We remark
that provided that eccentricity
damping limits eccentricities
to small values, the commensurabilities that are formed in the system
as a consequence of convergent migration depend
on the ratio of the  adopted  migration rate to the
local  orbital frequency,
with  commensurabilities of low order and low degree forming when this ratio is small.
\subsection{Numerical results}\label{Numresults}

\subsection{ A system with a 3:2 commensurability}\label{3:2}

            For the calculatioas presented in this section  we adopted masses
for the two planets and the central star that coincided
with those for the star and  two innermost planets of the GJ581 system.
Thus  the inner  planet was taken to have a mass  $m_1=1.94M_{\oplus}$
and to be in circular orbit at  $0.16au.$ 
The outer planet was taken to have a mass  
$m_i=15.64M_{oplus}$ and to be in a circular orbit at
$0.32au.$ Tests indicate that the results of  simulations of the type described
here do not depend on  the longitudes at which the planets are inserted
on such circular orbits. 
The central mass was $0.31M_{\odot}.$
The initial  semi-major axes were chosen to be larger than 
the corresponding ones in the GJ581 system so as to allow for some inward migration.
The disk migration and circularization rates
adopted were given by 
\begin{equation} t_{mig}=4.375\times 10^5\frac{M_{\oplus}}{m_i} yr.
 \hspace{2mm}{\rm and} \hspace{2mm}  t_{c,i}= 5\times 10^2\frac{M_{\oplus}}{m_i} yr.
\end{equation}
 However, they were only applied when the semi-major axis of a planet exceeded
 $0.041 au.$ This procedure results in the final semi-major axis of the
 outer planet to coincide with the second planet in the GJ581 system.
 The termination of disk migration could be regarded as either being due to
 entry into an inner cavity, or  simply removal of the disk.
 The migration rate was chosen so as to enable the planets to settle into a 3:2
commensurability through convergent migration. A very much slower rate would allow
trapping in a 2:1 commensurability, while a very much faster one would result in the system
passing through the 3:2 commensurability (see eg.  Papaloizou \& Szuskewicz 2010).
 We remark that although the specific parameters chosen correspond to
the GJ581 system, the aruments presented above indicate that the form of evolution
we find should be generic for two low mass planets attaining a first order commensurability
through convergent migration.
           
\begin{figure*}
 \includegraphics[width=\textwidth]{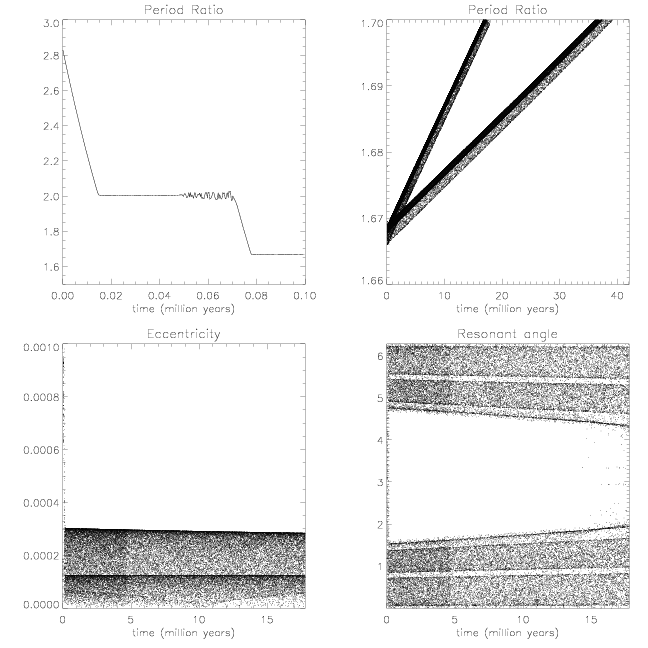}
\caption{The evolution of two planets that form a 5:3 commensurability.
The early evolution of the period ratio during  convergent migration
is shown in the upper left panel.  The upper right panel shows the
evolution of the period ratio under orbital circularization
after disk migration ceases. The uppermost curve is for $Q'=1.5$ and the
lower curve is for $Q'=3.$  The evolution of the eccentricity of the
outermost planet is plotted in the lower left panel. The evolution
of the resonant angle  $3\lambda_2-2\lambda_1-\varpi_1$
is plotted in the lower right panel.}
\label{fig:2}       
\end{figure*}

 The evolution of the system is illustrated in Fig. \ref {fig:1}.
The early evolution of the period ratio during  convergent migration
is shown in the upper left panel. It is seen that the system is trapped in a
2:1 commensurability for a while before escaping to be subsequently trapped
in a 3:2 commensurability. After about $2\times 10^4 yr.$ the forces from the disk
cease to act and the system evolves under tidal circularization.
   The upper right panel  of Fig. \ref{fig:1} shows the
evolution of the period ratio.  Results for  simulations with  $Q'=1.5$ and 
 $Q'=3$ are illustrated and compared to  analytic predictions
derived  from equation (\ref{sepeqt}) adapted to the cases on hand.
 Interestingly the  numerical results are in quite good agreement with 
what is expected  from the analytic discussion given in section \ref{p+1p}
which assumed a small libration amplitude  and which led to equation (\ref{sepeqt}),
even in regimes where the amplitude of libration of the resonant angle is quite large.
However,  the simulations show additional  sudden small jumps
in the period ratio which occur when the system passes through the 5:3 resonance.
This jump was  larger for the $Q'=1.5$ case than for the $Q'=3$ case.
 The evolution
of the resonant angle  $3\lambda_2-2\lambda_1-\varpi_1$
 for $Q'=1.5.$ shown in Fig. \ref{fig:1} indicates an increasing amplitude libration that
tends to break down near the end of the simulation when the period ratio $\sim 1.7$ as in GJ581.
But note that there is also a short temporary breakdown as the system passes through 5:3 resonance.
Note that the anlaytic treatment suggests the time for the period ratio to evolve from
1.5 to 1.7 to be  $\sim 5\times 10^6 yr.$ The simulation with $Q'=1.5.$
rather fortuitously agrees very well with this. The analytic prediction for  $Q'=3$ 
is $10^7 yr.$ while the simulations discussed in this and the next section
indicate $1.3\times 10^7 yr.$  Given the expectation  that evolution times are
$\propto Q',$ this indicates that values of $Q'$ as large as a few hundred
could have allowed the period ratio to move from $1.5$ to the present value within
the lifetime of the system.

\subsection{A system with a 5:3 commensurability}\label{5:3}

For the calculations presented in this section we adopted
  the same values for the  central mass and the planet masses 
 as in section \ref{3:2}. However, we adopted initial conditions,
 migration and circularization rates so as to enable the system to settle
into a 5:3 commensurability. Thus
the inner planet was started in circular orbit at $0.08au$ 
and the outer planet 
in a circular orbit at
$0.16au$ in this case. 
The disk migration and circularization rates
 adopted were given by
\begin{equation} t_{mig}=1.75\times 10^5\frac{M_{\oplus}}{m_i} yr.
 \hspace{2mm}{\rm and} \hspace{2mm}  t_{c,i}= 2\times 10^3\frac{M_{\oplus}}{m_i} yr.
\end{equation}
 Thus the convergent migration rate was two and a half times
  faster and the circularization rate four times slower than for the calculation
  in section  \ref{3:2}. However, they were applied in the same way. 
The faster migration rate and the slower eccentricity damping rate
 allows trapping in the 5:3 resonance.

The early evolution of the period ratio during  convergent migration
is shown in Fig. \ref{fig:2}.  
 It is seen that the system becomes trapped in a
2:1 commensurability  before escaping to form
a 5:3 commensurability. After about $8\times 10^4 yr.$ the forces from the disk
cease to act and the system evolves under tidal circularization.

 The upper right panel of Fig. \ref{fig:2}  shows the
evolution of the period ratio under orbital circularization
after disk migration ceases  for $Q'=1.5$ and 
 $Q'=3.$ Although the system started in a 5:3 commensurability,
  the evolution can be regarded as matching onto that illustrated in the
previous section which can be regarded as being driven by the $3:2$
comensurability. 
 This is also confirmed by the evolution
 the resonant angle  $3\lambda_2-2\lambda_1-\varpi_1$
also plotted in Fig.\ref{fig:2}. We also remark that
the time for the period ratio to move from $5/3$ to $1.7$ is
about twice as large for $Q'=3$ as for $Q'=1.5$ as expected.
However, these times are only approximately $1.8\times 10^6$ and $3.6\times 10^6 yr.$
respectvely indicating that values of $Q'$ up to  $10^3$ could be effective within
the lifetime of the system.

\begin{figure*}
 \includegraphics[width=\textwidth]{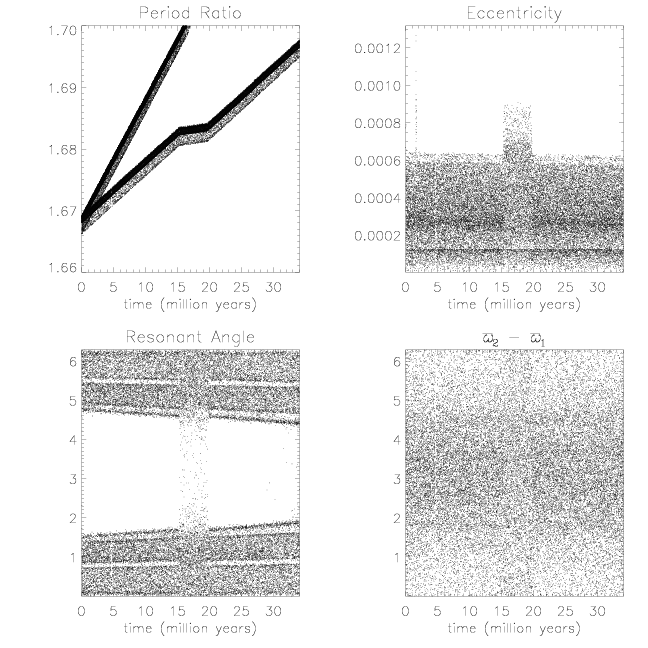}
\caption{The evolution of two planets illustrated in Fig. \ref{fig:3}
but with two additional outer planets added after the disk
migration phase as indicated in the text.
  The upper right panel shows the
evolution of the period ratio of the innermost  two planets under orbital circularization.
 The uppermost curve is for $Q'=1.5$ and the
lower curve is for $Q'=3.$  The evolution of the eccentricity of the
second innermost planet is plotted in the upper left panel. The evolution
of the resonant angle  $3\lambda_2-2\lambda_1-\varpi_1$
is plotted in the lower left panel. The behaviour of the  angle between
 the apsidal lines of the   orbits of the second innermost and innermost  planets is illustrated in the lower right panel.}
\label{fig:3}       
\end{figure*}

\subsection{Adding additional planets}\label{extraplanets}

  Here the effect of adding additional planets  to the simulation described above is investigated.
To do this we take the calculation of section \ref{5:3} at the point at which 
forces arising from the  disk cease to act.
Two additional planets of masses $5.36M_{\oplus}$ and $7.09M_{\oplus}$
are added in circular orbits with semi-major axes $0.07au$ and $0.22au$ respectively.
These correspond to the two outermost planets in the GJ581 system. We remark that
the eccentricities of these planets were determined to be consistent with zero
by Vogt et al. (2010).  
As before we considered runs for which $Q'=1.5$ and $Q'=3.$
In this case the same value of $Q'$ was adopted for each planet.

The results are plotted in Fig. \ref{fig:3}. The evolution in this case is for the most part
similar to that illustrated in Fig. \ref{fig:2} for two planets. In particular approximately
the same time is taken for the period ratio for the innermost pair of planets to move from $5/3$
to $ 1.7. $ However, a significant difference is that the evolution of the period
ratio slows down briefly between $1.5\times 10^7$ and $2.0\times 10^7 yr.$  in the simulation with $Q'=3.$
During this time the eccentricity of the second innermost planet is increased.
Although the reasons for this are unclear,  it is associated with an interaction between
the second and third innermost planets. The innermost planet continues to move inwards
but the angular momentum ends up being transferred to the third rather than the second innermost planet.
There does not seem to be any clear resonance associated with this. However, we comment
that in a many planet system like this, we could consider a tension between possible interacting pairs.
The second and third planets would separate on account of tidal circularization if the
innermost planet were absent. Similarly the innermost pair can couple as in section \ref{5:3}.
In some circumstances, dependent on their masses, orbital parameters, and values of  $Q'$ etc.,
different interacting pairs may have varying levels of importance in the simulation.
This requires a more detailed study than we have been able to perform at this preliminary stage
that will be the subject of future work.

\subsection{A system with a 3:1 commensurability}\label{3:1}
Finally we   describe a situation in which a $3:1$ commensurability
could be formed under convergent migration and then subsequently maintained.
The parameters of this simulation were chosen  to lead to a  separation of pairs similar to  the third and fourth 
innermost  planets in the  HD 10180 system for  which  there may have been  a past proximity to a 3:1 commensurability. 

\begin{figure*}
 \includegraphics[width=\textwidth]{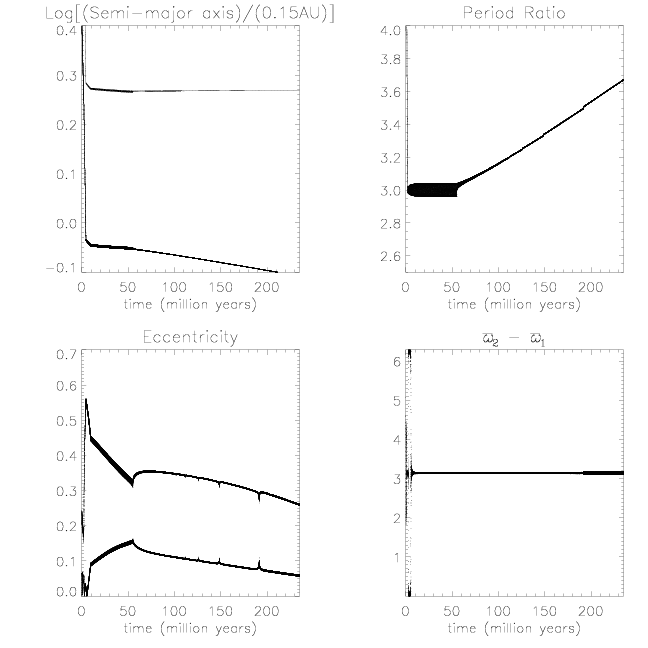}
\caption{The evolution of two planets in a 3:1 commensurability is illustrated.
The left uppermost panel shows the evolution of the semi-major axes of the two planets.
The initial period of  disk migration is short $< \sim 4.5\times10^{6} yr.$ The subsequent evolution
is driven by tidal circularization with $Q'=100$  and the gravitational interaction between the planets.
The upper left panel shows the evolution of the period ratio.  This remains 3:1 for  some time after
disk migration has ceased before finally increasing  as the planets separate.
The lower left panel shows the evolution of the eccentricities of the two planets, the uppermost curve
corresponding to the inner planet. The lower right panel  shows the  angle between the
apsidal lines of the  orbits of the  outer and inner  planets, which ultimately remains close to $\pi.$ }
\label{fig:4} 
\end{figure*}

In this case the central mass was taken to be $1M_{\oplus}.$
The inner planet mass was taken to be $11.73M_{\oplus}$ and the outer planet taken 
to be  $25.07M_{\oplus}.$
Their initial semi-major axes were 
$0.387au$ and $1.2au$ respecively. The outer planet was started in circular orbit
while the inner planer was started at apocentre with an eccentricity $e=0.24.$
   The disk migration and circularization rates
  adopted were given by
\begin{equation} t_{mig}=  t_{c,i}= 1.4\times 10^8\frac{M_{\oplus}}{m_i} yr.
\end{equation}
These were applied only when the planets semi-major axes exceeded $0.29au.$
Note that this migration rate is very much lower than the previous cases
so as to enable  trapping in the $3:1$ resonance. The ecentricity damping rate is taken to be
equal to the migration rate so that the eccentricities do not damp too quickly
so enabling the 3:1 resonance to persist. Rates like these are not readily produced in calculations
of disk planet interactions for which the planets are fully embedded. They may be possible if the planets
are located within a wide cavity. However, this aspect remains to be investigated.
Here we simply adopt these rates and explore their consequences.
Because of the larger planetary masses  and larger orbital eccentricities
in this run, it was possible to consider larger values of $Q'.$ We  adopted $Q'=100.$

The evolution of the two planets in this simulation is illustrated in Fig. \ref{fig:4}.
The planets undergo convergent migration and attain a 3:1 resonance. The eccentricity of the inner planet grows
up to $\sim 0.56.$ The growth ceases after $\sim 4\times 10^6 yr.$ when effects  arising from the disk
cease to act. After this time the planets evolve under tidal circularization.
For about $50$ million years the commensurability  is maintained while the eccentricty of the inner planet
decreases and that of the outer one increases. In this process angular momentum is transferred
to the inner planet. However, this form of evolution cannot be maintained and it reverts to the situation where
the planets separate in  semi-major axis as descibed in section \ref{simple1} while the eccentricities decrease.
The period ratio secularly increases  while the angle between the apsidal lines  of the orbits  of  the
outer and inner planets   remains close to $\pi.$

 Interestingly at least five resonance passages were seen during this  later evolutionary stage.
 As the period ratio increased, these corresponded to the 19:6, 16:5, 13:4, 10:3 and 7:2
 resonances.  They are manifested as local blips in the eccentricity evolution
of  both planets as shown in the lower left panel of 
Fig. \ref{fig:4}.
The resonance passages are of decreasing order with increasing time and so 
the consequent changes induced in the  planetary eccentricities
increase in magnitude.
The fact high order resonances  such as $19:6$ were manifest in this run is because
of the relatively  high eccentricities, in particular
of  $\sim 0.3$  for  the inner planet. 

\section{Discussion}\label{Discuss}
In this paper we have studied systems of close orbiting planets evolving under the influence
of tidal circularization. We considered  the situation where  the system evolved under the influence
of disk tides to form a commensurability. After the disk tides ceased to operate, either because of
entry into an inner cavity, or because of  loss of the disk,  the operation of  tidal circularization
caused increasing  departure from any close  commensurability as time progressed.

In  section\ref{simple1} we pointed out that  a system of planets in near circular orbits
is expected to separate on average as energy is dissipated while angular momentum is conserved.
This is also expected in the very similar situation of an accretion disk evolving under a viscosity
(see Lynden-Bell \& Pringle 1974). In the simplest case of two planets, 
this inevitable increasing physical 
separation has to lead to the  increasing departure from any  initial commensurability.
 
 In sections \ref{coords}, \ref{disk tides}  and  \ref{p+1p}, we developed a formalism 
  that could be adapted study the evolution
of two planets near to a first order commensurability  under the influence of tidal circularization.
 This was then applied to a system with a  tight commensurability in section  \ref{tight}.
 An expression for the  departure  from commensurability,  indicating this to be $\propto t^{1/3},$ 
  was given  (see equation (\ref{sepeqt})).
 The discussion was then extended to the situation when
the two planets were  not necessarily in  a close commensurability  in section \ref{loose}.
The orbital evolution of the planet in that case, leading to
a neighbouring  commensurability, was then considered in section \ref{orbevol}.

In order to confirm the analytic modeling,  numerical simulations 
were under taken in section \ref{Numsim}.
We were able to set up systems of low mass planets in varying commensurabilities,
depending on the strengths 
of the disk tides leading to orbital migration and circularization,
with weaker tides in general leading to more widely
separated commensurabilities.  We focused on a two planet system which had the same parameters 
as the innermost two planets as the GJ581 system in section \ref{3:2}.  This formed 
 a 3:2 commensurability
which then evolved under orbital circularization.
This model system attained the period ratio of the actual system after $\sim 10^8 yr.$ when $Q'\sim 1.$
Simple extrapolation thus indicates that tidal evolution could have moved the system
to the present  period ratio of $1.7$  from a 3:2 commensurability if $Q' \sim 100.$
Similarly  the situation  when the system initially attained 
 a 5:3 commensurability was studied in section \ref{5:3}.
In this case the evolution quickly adapted to evolve as
 for the case with the initial 3:2 commensurability
when that had reached the same period ratio.  However, a larger value $Q'$ would
suffice to  cause  the period ratio
to move from  5:3  to the observed one within a given life time.
The effect of  adding  the additional planets in the GJ581 system was considered in section \ref{extraplanets}.
In that case over the  long term  the extra planets did not greatly affect the evolution.
However,   for a brief period the third planet moved outwards taking up the angular momentum
of the innermost planet rather than the second, with the consequence that the period separation rate
for the innermost pair was slowed. Thus a pair of planets may not always  evolve independently
of others in the system, a feature  that requires further study. 
 
 Finally the evolution of a  system that formed  a 3:1 commensurability 
was considered in section \ref{3:1}.
 The model system adopted had similar parameters  to the third and fourth innermost planets
 in the  HD 10180 system. This case required slow disk migration and weak circularization
with the result that the resonance involved high eccentricities.  Because of these the commensurability could 
be maintained for a while under orbital circularization. However,  eventually the system increasingly   departed
from it as in the other cases.
Finally all of our results indicate that  if $Q'   <\sim 100,$  commensurabilities would have been significantly affected
by tidal effects related to orbital circularization. 
 Thus the survival of close  commensurabilities in   observed systems
may be indicative of the presence of large $Q'$ values, a feature which in turn may be related to
the internal structure of the planets involved.


\begin{acknowledgements}
This work was supported by the Science and
Technology Facilities Council [grant number ST/G002584/1].
\end{acknowledgements}

\end{document}